\documentstyle[12pt,slipl]{article}
\lefthead{AFRAIMOVICH ET AL.}
\righthead{MAGNETOSPHERIC DISTURBANCES, AND THE GPS OPERATION}
\received{}
\revised{}
\accepted{}

\authoraddr{E.~L. Afraimovich,
Institute of Solar-Terrestrial Physics SD RAS,
P.~O.~box~4026, Irkutsk, 664033, Russia,
fax: +7 3952 462557; e-mail:~afra@iszf.irk.ru}

\begin{document}

\title{Magnetospheric disturbances, and the GPS operation}

\author{Afraimovich~E.~L., O.~S.~Lesyuta, and I.~I.~Ushakov\\
Institute of Solar-Terrestrial Physics SD RAS,\\
P.~O.~box~4026, Irkutsk, 664033, Russia\\
fax: +7 3952 462557; e-mail:~afra@iszf.irk.ru}

\date{}

\begin{abstract}
We have investigated a dependence of the relative density of
phase slips in the GPS navigation system on the disturbance level
of the Earth's magnetosphere. The study is based on using
Internet-available selected data from the global GPS network,
with the simultaneously handled number of receiving stations
ranging from 160 to 323. The analysis used four days from the
period 1999--2000, with the values of the geomagnetic field
disturbance index $Dst$ from 0 to -300 nT. During strong magnetic
storms, the relative density of phase slips on mid latitudes
exceeds the one for magnetically quiet days by one-two orders of
magnitude as a minimum, and reaches a few and (for some of the
GPS satellites) even ten percent of the total density of
observations. Furthermore, the level of phase slips for the GPS
satellites located on the sunward side of the Earth was by a
factor of 5-10 larger compared with the opposite side of the
Earth. The high positive correlation of an increase in the
density of phase slips and the intensity of ionospheric
irregularities during geomagnetic disturbances as detected in
this study points to the fact that the increase is slips is
caused by the scattering of the GPS signal from ionospheric
irregularities.
\end{abstract}

\begin{article}

\section{Introduction}

The satellite navigation GPS system has become a powerful factor
of scientific and technological progress worldwide, and enjoys
wide use in a great variety of human activity. In this
connection, much attention is given to continuous perfection of
the GPS system and to the widening of the scope of its
application for solving the navigation problems themselves, as
well as for developing higher-precision systems for time and
accuracy determinations. Even greater capabilities are expected
in the near future through the combined use of the GPS with a
similar Russian system (GLONASS).

Recently the GPS system has also gained wide-spread acceptance in
research in the field of geodynamics, in the physics of the
Earth's atmosphere, ionosphere and plasmasphere, etc. [{\it
Davies and Hartmann}, 1997; {\it Klobuchar}, 1997].
Investigations of this kind are not only of purely scientific
interest but are also important for perfection of the GPS system
itself. To address these problems, a global network of receiving
GPS stations was set up, which consisted, by November 2000, of no
less than 757 points, the data from which are placed on the
Internet.

Using two-frequency multichannel receivers of the global
navigation GPS system, at almost any point on the globe and at
any time simultaneously at two coherently-coupled frequencies
$f_1=1575{.}42$ MHz and $f_2=1227{.}60$ MHz, highly accurate
measurements of the group and phase delays are being underway
along the line of sight (LOS) between the receiver on the ground
and the transmitters on-board the GPS system satellites which are
in the zone of reception.

These data, converted to values of total electron content (TEC),
are of considerable current use in the study of the regular
ionosphere and of disturbances of natural and technogenic origins
(solar eclipses, flares, earthquakes, volcanoes, strong
thunderstorms, auroral heating; nuclear explosions, chemical
explosion events, launches of rockets). We do not cite here the
relevant references for reasons of space, which account for
hundreds of publications to date.

The study of deep, fast variations in TEC caused by a strong
scattering of satellite signals from intense small-scale
irregularities of the ionospheric $F_2$-layer at equatorial and
polar latitudes has a special place among ionospheric
investigations based on using satellite (including GPS) signals
[{\it Aarons et al}., 1996, 1997; {\it Klobuchar}, 1997; {\it Pi
et al}., 1997; {\it Aarons and Lin}, 1999]. The interest to this
problem as regards the practical implementation is explained by
the fact that as a result of such a scattering, the signal
undergoes deep amplitude fadings, which leads to a phase slip at
the GPS working frequencies.

To achieve a more effective detection of disturbances in the
near-terrestrial space environment, we have developed a new
technology of a global detector GLOBDET, and a relevant software
which makes it possible to automate the acquisition, filtering and
pretreatment process of the GPS data received via the Internet [{\it
Afraimovich}, 2000b]. This technology is being used to detect, on a
global and regional scales, ionospheric effects of strong magnetic
storms [{\it Afraimovich et al}., 2000a], solar flares [{\it
Afraimovich}, 2000b, 2000c], solar eclipses [{\it Afraimovich et al}.,
1998], launches of rockets [{\it Afraimovich et al}., 2000d],
earthquakes, etc.

In this paper we have used an earlier GLOBDET technology in a global
analysis of the relative density of phase slips in the GPS system during
disturbances of the near-terrestrial space environment. The
experimental geometry and general information about the data base
used are presented in Section~2. The determination of
the relative density of phase slips, and the method of processing the
data available from the Internet are briefly outlined in
Section~3. Section~4 describes the
results obtained for magnetically quiet and disturbed conditions. Results
are discussed in Section~5.

\section{Experimental geometry and general information about the data
base used}

This study is based on using the data from a global network of
GPS receiving stations available from the Internet. For a number
of reasons, slightly differing sets of GPS stations were chosen
for the various events under investigation; however, the
experimental geometry for all events was virtually identical. The
analysis used a set of stations (from 160 to 323) with a
relatively even distribution across the globe. For reasons of
space, we do not give here the stations coordinates. This
information may be obtained from
http://lox.ucsd.edu/cgi-bin/allCoords.cgi?.

The set of stations, which we selected out of the part of the
global GPS network available to us, covers rather densely North
America and Europe; Asia has much poorer coverage. The number of
GPS stations in the Pacific and Atlantic oceans is even smaller.
However, such coverage over the globe is already presently
sufficient for a global detection of disturbances with spatial
accumulation unavailable before. Thus, in the western hemisphere,
the corresponding number of stations can, already today, reach at
least 500, and the number of beams to the satellites no less than
2000-3000.

The analysis involved four days of the period 1999-2000, with the
values of the geomagnetic field disturbance index $Dst$ ranging
from 0 to -300 nT and $Kp$ from 2 to 9. These events are
summarized in Table~1.

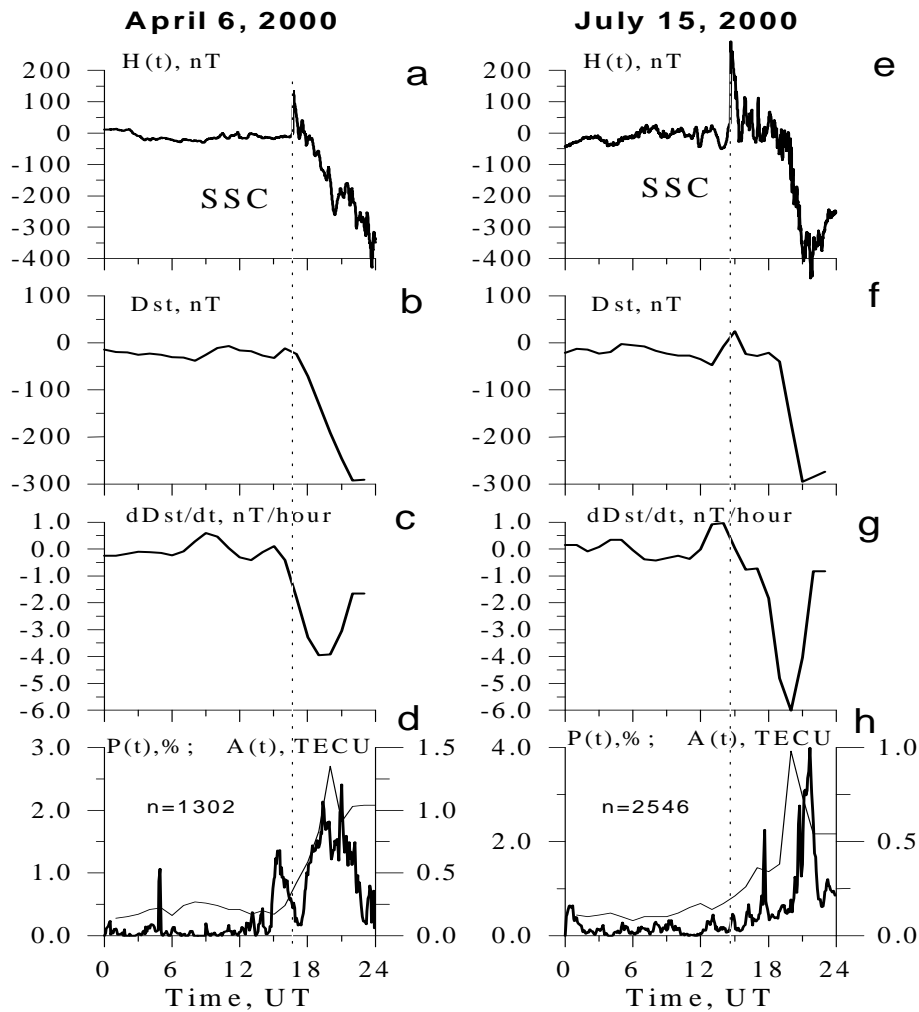
Figure~1 presents the measured variations of the $H$-component of
the geomagnetic field at station Irkutsk ($52{.}2^\circ N$;
$104{.}3^\circ E$ --a, e), and $Dst$ (b, f) during major magnetic
storms on April 6, and July 15, 2000; a correlative analysis of
the data is made in Sections~4.

The statistic of the data used in this paper for each of the days
under examination is characterized by the information in Table~1
about the number of stations used $m$, and in Tables~2, 3 about
the total number $n$ of satellite GPS passes (LOS). The total
amount of data exceeds $10^{7}$ 30-s observations.

\section{The method of processing the data from the Internet}

The purpose of a preprocessing of the GPS data in this paper is
to obtain slip density estimates in measuring the phase
difference $L1-L2$, and slips of phase measurement at the
fundamental frequency $L1$. Ascertaining the cause of the
increase in slip density was also greatly facilitated by
estimating the TEC variation intensity for the same stations and
time intervals.

The GPS technology provides the means of estimating TEC
variations on the basis of phase measurements of TEC $I$ in each
of the spaced two-frequency GPS receivers using the formula
[{\it Afraimovich et al.}, 1998]

\begin{equation}
\label{Spe-eq-01}
I_o=\frac{1}{40{.}308}\frac{f^2_1f^2_2}{f^2_1-f^2_2}
                           [(L_1\lambda_1-L_2\lambda_2)+const+nL]
\end{equation}

\hspace{-0.5 cm} where $L_1\lambda_1$ and $L_2\lambda_2$~ are
phase path increments of the radio signal, caused by the phase
delay in the ionosphere (m); $L_1$ and $L_2$~ are the number of
full phase rotations, and $\lambda_1$ and $\lambda_2$ are the
wavelengths (m) for the frequencies $f_1$ and $f_2$,
respectively; const~ is some unknown initial phase path (m); and
$nL$~ is the error in determination of the phase path (m).

Phase measurements in the GPS can be made with a high degree of
accuracy corresponding to the error of TEC determination of at
least $10^{14}$~m${}^{-2}$ when averaged on a 30-second time
interval, with some uncertainty of the initial value of TEC,
however [{\it Hofmann-Wellenhof et al}., 1992]. This makes possible
detecting ionization irregularities and wave processes in the
ionosphere over a wide range of amplitudes (up to $10^{-4}$ of
the diurnal TEC variation) and periods (from 24 hours to 5 min).
The TECU (Total Electron Content Units), which is equal to
$10^{16}$ m${}^{-2}$ and is commonly accepted in the literature,
will be used throughout the text.

Primary data include series of "oblique" values of TEC $I_o(t)$, as
well as the corresponding series of elevations $\theta(t)$ and
azimuths $\alpha(t)$ along LOS to the satellite calculated using
our developed CONVTEC program which converts the GPS system
standard RINEX-files on the Internet [{\it Gurtner}, 1993].

\subsection{The relative density of difference phase $L1-L2$ slips and
of phase $L1$ slips}

We hold fixed a slip of the phase difference $L1-L2$ in the case
where the modulus of the TEC increment for a time interval of 30
s (which is a standard one for the GPS data placed on the
Internet) that is calculated by formula (1), exceeds the
specified threshold of order, for example, 100-200 TECU. A slip
of phase $L1$ is also fixed in a similar manner but with a much
larger threshold and with due regard for the time varying
distance to the satellite.

As a result of a pretreatment of the RINEX-files, we have the number
$S$ of phase slips within a single selected time interval $dT$=5 min,
as well as the corresponding number $M$ of observations that is
required for normalizing the data. Our choice of such an interval was
dictated by the need to reduce the amount of the data analyzed without
decreasing the time resolution that is required for the analysis (a
standard time step for the RINEX-files equal to 30 s would require a
larger memory capacity).

These data for each of the GPS satellites were then averaged for
all the stations selected in order to infer the mean density of
observations $M(t)$ and the mean density of phase slips $S(t)$.
In the middle of the observed satellite pass, the density of
observations $M(t)$ averages 10$\pm$1 (30-s counts); at the
beginning and end of the pass it can decrease because the time
intervals of observation of a given satellite at elevations
larger than that specified do not coincide at different stations.
Subsequently, we calculated the mean relative density of phase
slips $P(t)$=$S(t)$/$M(t)$, $\%$. Furthermore, the daily mean
value of the relative number of phase slips $\langle P\rangle$
that was averaged over all GPS satellites and stations was useful
for our analysis.

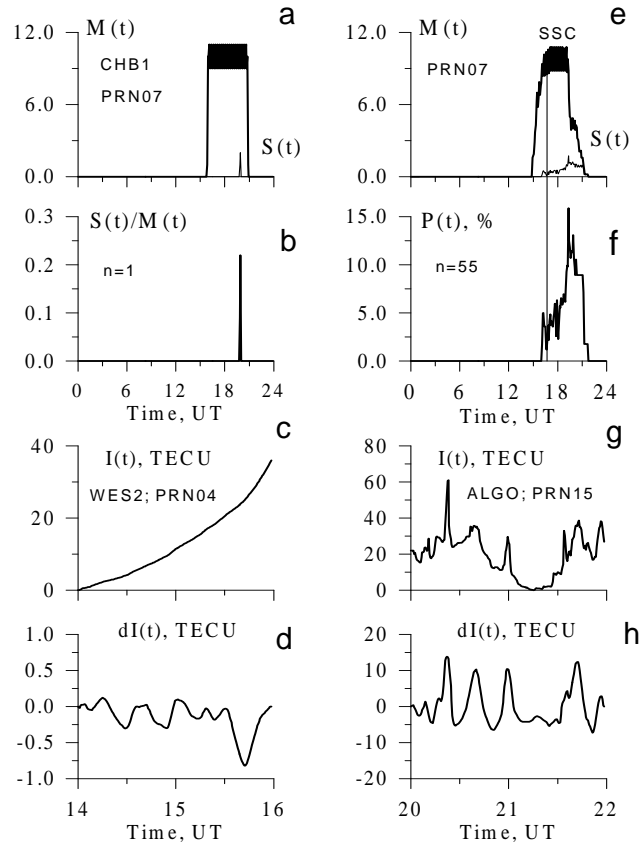
Figure~2 (at the left) gives an example of the observation density
$M(t)$ --a, heavy line; the slip density $S(t)$--a, thin line; b - ratio
$S(t)/M(t)$ for one of the PRN07 passes as recorded at station CHB1
(the geographic coordinates are $45.6^\circ$N, $275.5^\circ$E) during
the magnetic storm of April 6, 2000. Dependencies $M(t)$, $S(t)$ and
of the relative slip density $P(t)$ for PRN07, averaged over the
mid-latitude stations of North America as a function of universal time
(UT), are plotted in panels (e,f). The SSC (sudden storm
commencement) time 16{:}42 UT is shown in panels e) and f) by a
vertical bar. The number $n$ of satellite PRN07 passes used to carry
out an averaging is marked in panel f. In this paper we are using the
term PRN (pseudo random noise) to designate the satellite number [{\it
Hofmann-Wellenhof et al}., 1992].

As would be expected the mean observation density $M(t)$ for a
single satellite exhibits a diurnal variation that is determined
by the satellite's orbit, and varies over the range from 0 to 8.

\subsection{Estimation of the TEC variation intensity}

We have used the series $I_o(t)$, containing neither slips of the
phase difference $L1-L2$ nor gaps of counts, to estimate the TEC
variation intensity for the same sets of stations and time
intervals as used in estimating the phase slip density.

Series of the values of elevations $\theta(t)$ and azimuths
$\alpha(t)$ of the beam to the satellite were used to determine
the coordinates of subionospheric points, and to convert the
"oblique" TEC $I_{o}(t)$ to the corresponding value of the
"vertical" TEC by employing the technique reported by
[{\it Klobuchar,} 1986]

\begin{equation}
\label{SPE-eq-02}
I = I_o cos \left[arcsin\left(\frac{R_z}{R_z + h_{max}}cos\theta\right)
\right]
\end{equation}

where $R_{z}$ is the Earth's radius, and $h_{max}$=300 km is the height
of the $F_2$-layer maximum.

To exclude the variations of the regular ionosphere, as well as
trends introduced by the motion of the satellite, we employ the
procedure of removing the linear trend by preliminarily smoothing
the initial series with a selected time window of a duration of
about 60 min. In a subsequent treatment, we use the standard
deviation of the TEC variations $dI(t)$, thus filtered, as an
estimate of the TEC variation intensity $A$ (see Section 4).

Figure~2c gives an example of a typical weakly disturbed variation in
"oblique" TEC $I_o(t)$ for station WES2 (satellite number PRN04) on
July 15, 2000 for the time interval 14{:}00-16{:}00 UT, preceding the
onset of a geomagnetic disturbance near the WES2 station
($42.6^\circ$N, $288.5^\circ$E). For this same series, Figure~2d
presents the $dI(t)$ variations that were filtered out from the $I_o(t)$
series by removing the trend with a 60-min window (rms of $dI(t)$ is
smaller then 0.2 TECU).

Strong variations in TEC variation intensity occurred near
station WES2 literally within 6 hours. Figure~2g and Figure~2h
presents the dependencies $I(t)$ and $dI(t)$ for station ALGO
($46^\circ$N, $282^\circ$E), July 15, 2000, for the time
interval 20{:}00-22{:}00 UT (PRN15). As is evident from the
figure, the TEC variations increased in intensity at least by a
factor of 40-50 when compared with the time interval
14{:}00--16{:}00 UT (Figure~2c and Figure~2d).

\subsection{Conditions and limitations of a data processing}

Slips of phase measurements can be caused by reception conditions
for the signal in the neighborhood of the receiver (interference
from thunderstorms, radiointerferences), which is particularly
pronounced at low elevations $\theta$. To exclude the influence
of the signal reception conditions, in this paper we used only
observations with satellite elevations $\theta$ larger than
$30^\circ$.

Another possible reasons for the phase slips, as has been pointed
out in the Introduction, is due to deep, fast changes in TEC
because of a strong scattering of satellite signals from intense
small-scale irregularities of the ionospheric $F_2$-layer at
equatorial and polar latitudes [{\it Aarons et al}., 1996, 1997;
{\it Klobuchar}, 1997; {\it Pi et al}., 1997; {\it Aarons and
Lin}, 1999].

However, since we are using a global averaging of the number of phase
slips for all beams and stations, as a consequence of the uneven
distribution of stations the proportion of mid-latitude stations of North
America and, to a lesser extent, of Europe is predominant (see above).
At the same time the number of stations in the polar region of the
northern hemisphere and in the equatorial zone was found to be quite
sufficient for a comparative analysis. To compare the results, we chose
3 latitude ranges: high latitudes $50-80^\circ$N; mid-latitudes
$30-50^\circ$N; and equatorial zone $30^\circ$S--$30^\circ$N.

We selected also the data according to the types of two-frequency
receivers, with which the GPS global network sites are equipped
(the relevant information is contained in the initial RINEX
format).

\section{Results derived from analyzing the relative density of phase
slips}

\subsection{Magnetically quiet days}

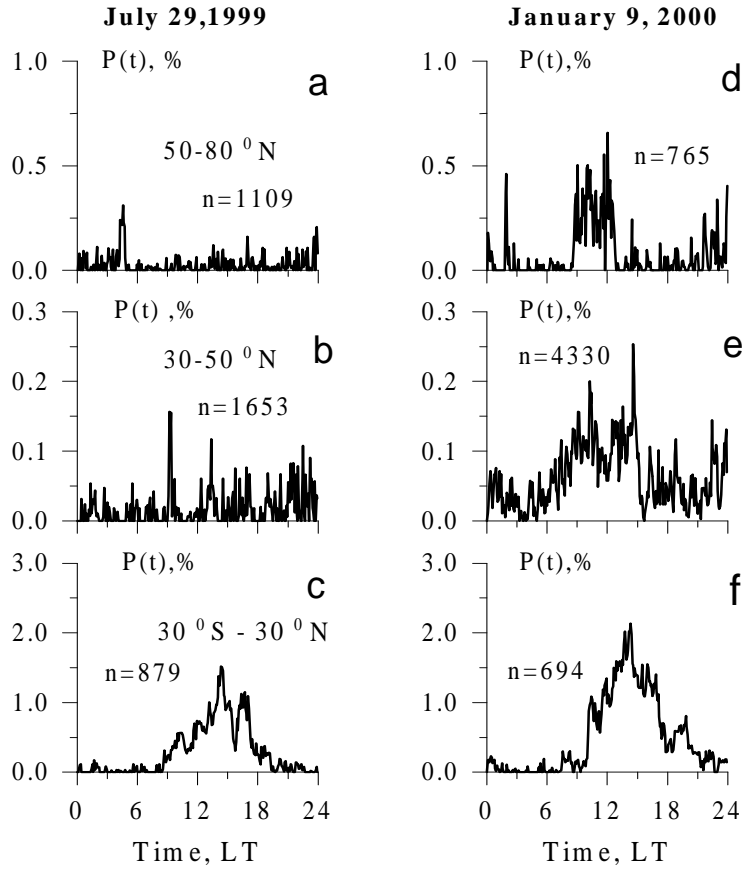
Figure~3 plots the local time LT-dependence of the relative mean slip
density $P(t)$ obtained by averaging the data from all satellites in the
latitude range $0-360^\circ$E irrespective of the type of GPS receivers
for the magnetically quiet days of July 29, 1999 (at the left) and
January 9, 2000 (at the right). The local time for each GPS station was
calculated, based on the value of its geographic longitude. The number
$n$ of satellite passes used to carry out an averaging is marked in all
panels.

As is evident from Figure~3b, the phase slips on a magnetically quiet
day at mid latitudes have a sporadic character. The daily mean value of
the relative density of phase slips $\langle P\rangle$, averaged over all
GPS satellites and stations, was 0.017 $\%$ for the magnetically quiet
day of July 29, 1999 (the line 2 in the Table~2). Similar data were also
obtained for high latitudes (Figure~3a and the line 1 in the Table 2).

In the equatorial zone, however, even on a magnetically quiet
day, the density of phase slips exceeds the latitudinally mean
value of $P(t)$ at least by a factor of 15, and shows a strongly
pronounced LT-dependence, with a maximum value of 1.52 $\%$
(Figure~3c and the line 3 in the Table 2).

For the other magnetically quiet day of January 9, 2000, however, the
mean value of $\langle P\rangle$ at mid-latitudes was already larger
0{.}06 $\%$ (Figure~3e and the line 2 in the Table~2). For the diurnal
$P(t)$-dependence on January 9, 2000, one can point out the
irregularity of the mean density of phase slips as a function of local
time LT.

\subsection{Magnetic storms of April 6 and July 15, 2000}

A totally different picture was observed on April 6, 2000 during
a strong magnetic storm with a well-defined SSC.

Figure~1d presents the variations of the UT-dependence of the
relative mean slip density $P(t)$ obtained for the territory of
North America at mid-latitudes $30-50^\circ$N by averaging the
data from all satellites.
In this case, with the purpose of achieving a clearer detection
of the effect of the magnetic storm SSC influence on the
$P(t)$-dependence, we chose only those GPS stations which were on
the dayside of the Earth at the SSC time (North America region).
Noteworthy is a well-defined effect of an increase in
the density of phase slips that occurred after SSC.

A maximum mean slip density $P_{max}$=2.4 $\%$ is attained 3-4
hours after an SSC. For a separate satellite, PRN07, $P_{max}$ can
even exceed 10$\%$ (see Figure~2f). The same values, averaged over
all observed satellites and mid-latitude stations ($0-360^\circ$E) but as
a function of local time LT, are plotted in Figure~4b).

First of all, it should be noted that the relative density of
phase slips $P(t)$ exceeds that for magnetically quiet days by
one (when compared with January 9, 2000) or even two (when
compared with July 29, 1999) orders of magnitude, and reaches a
few and (for some of the GPS satellites) even ten percent of the
total observation density (Figure~2f). The mean value of $\langle
P \rangle$ for this storm is 0.67 $\%$ (the line 5 in the
Table~2), which is by a factor of 40 larger than that of $\langle
P\rangle$ for July 29, 1000, and by a factor of 10 larger than
that for January 9, 2000.

It was also found that the averaged (over all satellites) level
of phase slips for the GPS satellites on the subsolar side of the
Earth is by a factor of 10 larger than that on the opposite side
of the Earth (Figure~4b).

Similar dependencies with a maximum slip density $P_{max}$=3.37
$\%$ and a sharply pronounced diurnal dependence were also
obtained for equatorial latitudes (Figure~4c and the line 6 in
the Table 2). On the other hand, although the high latitudes show
a 10-fold increase of $\langle P \rangle$ as against a
magnetically quiet day, no LT-dependence is observed (Figure~4a
and the line 4 in the Table 2).

A similar result confirming all of the above-mentioned features
of the April 6, 2000 storm was also obtained for the other
magnetic storm of July 15, 2000 (see the measurements at magnetic
observatory Irkutsk in Figure~1e, the universal time UT
dependence in Figure~1h and the local time LT dependencies of the
relative mean density of phase slips $P(t)$ obtained by averaging
the data from all GPS satellites, in Figure~4d, e, f.

The mean value of $\langle P\rangle$ for this storm at mid-latitudes is
0{.}34 $\%$ (the line 5 in the Table~2), which is also in appreciable
excess of the level of phase slips for magnetically quiet days. The
effect of an increase in the density of phase slips after SSC is clearly
pronounced for this storm as well (Figure~1h; see below).

\subsection{Correlation of the increase in slip density and TEC
variation intensity}

It is known that equatorial latitudes are characterized by strong
scintillations of the transionospheric signal caused by the
scattering from $F_2$-region ionization irregularities [{\it
Aarons et al}., 1996, 1997; {\it Klobuchar}, 1997; {\it Pi et
al}., 1997; {\it Aarons and Lin}, 1999]. This is in reasonably
good agreement with our data on the diurnal dependence of the
phase slip density at the equatorial chain of stations
(Figures~3c, 4c).

Since during the active phase of the magnetic storm the
mid-latitude ionosphere becomes increasingly inhomogeneous, it
might be anticipated that a similar mechanism is able to cause
appreciable scintillations of the GPS signal at mid-latitudes as
well. To verify this hypothesis, we determined the dependencies
$A(t)$ of the TEC variation intensity obtained for the same set
of stations as in the case of $P(t)$ (see Section~3).

The dependencies $A(t)$ as a function of UT (Figure~1d, h - thin
line) and LT (Figure~4b, e-- thin line) presented below by
averaging (over all GPS satellites and stations) the standard
deviation of the variations $dI(t)$ for time intervals of 2.5
hours with a shift of 1 hour. Thus, 24 counts of the dependence
$A(t)$ are obtained for a 24-hour period with due regard for the
data from the preceding and next days.

In Figure~1h, the thin line represents the dependence $A(t)$ of
the TEC variation intensity obtained for all satellites and for
the territory of North America at mid-latitudes $30-50^\circ$N
during the magnetic storm of July 15, 2000. As is apparent from
this figure, the dependence $A(t)$ correlates quite well with the
UT-dependence of the relative mean slip density $P(t)$ calculated
from the same set of stations as in the case of $A(t)$.

A similar result on the UT-dependence was also obtained for a major
magnetic storm of April 6, 2000 (Figure~1d, thin line). A correlation of
the increase in slip density and in TEC variation intensity is shown as
clearly by the LT-dependencies $A(t)$ for the magnetic storms of
April 6 and July 15, 2000, presented in Figure~4b, e (thin line).

It was found that an increase in the level of geomagnetic
activity is accompanied by an increase in total intensity of
$A(t)$; however, it correlates not with the absolute level of
$Dst$, but with the value of the time derivative of $Dst$ (a
maximum correlation coefficient reaches -0{.}94 -- Figure~1c, g).
The derivative $d(Dst)/dt$ was obtained from the dependence
$Dst(t)$ (Figure~1b, f) that was smoothed with a 7-hour time
window.

This result is in reasonably good agreement with the conclusions
drawn in [{\it Ho et al.,}1998]; [{\it Afraimovich et al.}, 2000a].

\subsection{ The dependence of the slip density of phase
measurements $L1-L2$ and $L1$ on the type of GPS receivers }

The sample statistic of phase slips for the main types of
two-frequency receivers (ASHTECH, TRIMBLE, AOA), installed at the
global GPS network sites, is presented in Table 3. An analysis of
slips in measuring the phase difference $L1-L2$ and of the phase
$L1$ was carried out for the magnetically quiet day of July 29,
1999, and for the magnetic storms of April 6 and July 15, 2000.
Lines 1, 5, 9 and 13 reproduce the data derived from analyzing
the slips obtained for all stations irrespective of the type of
receivers, and given in lines 2, 3 and 5 of Table 2.

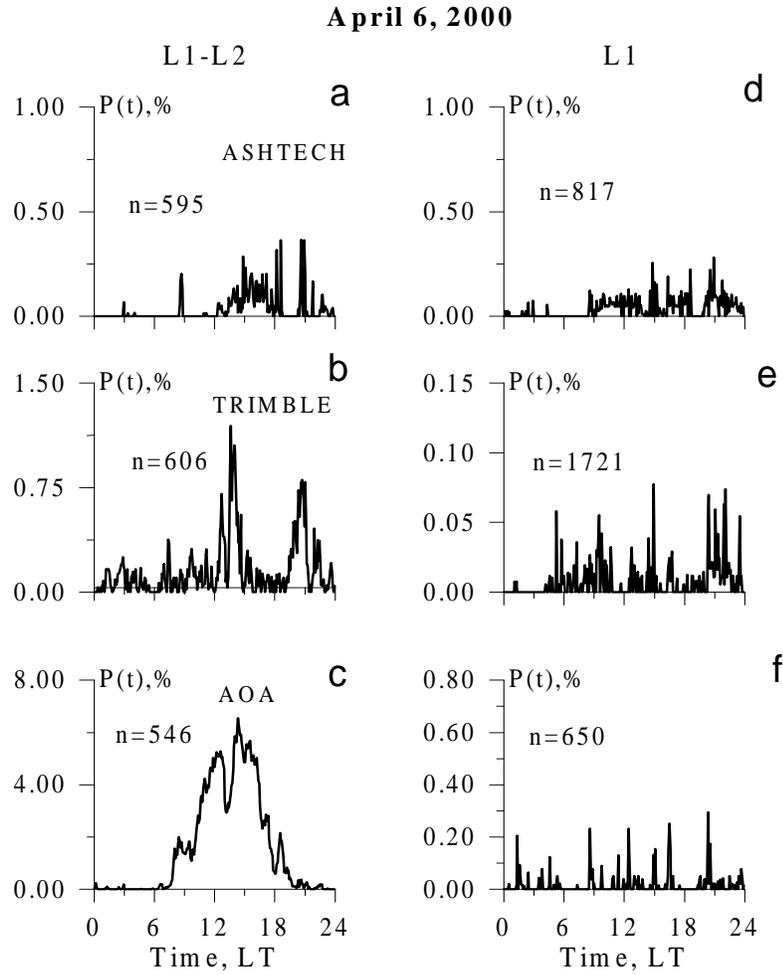
Figure~5 plots the LT-dependencies of the relative mean slip
density $P(t)$ of phase measurements of $L1-L2$ (at the left) and phase
measurements of $L1$ only (at the right) obtained by averaging the data
from all satellites in the longitude range $0-360^\circ$E at the
mid-latitudes $30-50^\circ$N for major magnetic storm on April 6,
2000.

The data from Table 3 and Figure~5a suggest that for the ASHTECH
receivers, the slip density of phase measurements at two
frequencies $L1-L2$ is by a factor of 5-20 smaller than that for
the other types of receivers. These slips have a sporadic
character, and show no clearly pronounced diurnal dependence.
These estimates are only slightly exceeded by the slip density
for the TRIMBLE receivers (Figure~5b).

The AOA receivers are the most susceptible to slips of $L1-L2$
measurements (Figure~5c). The mean and maximum values of $P(t)$
exceed the respective values for the ASHTECH receivers during the
magnetic storm of April 6, 2000, at least by a factor of 20-50.
Furthermore, the LT-dependence is the most pronounced. It should
also be noted that even on the magnetically quiet day of July 29,
1999, the level of slips for this receiver in the equatorial zone
is an order of magnitude higher when compared with mid-latitudes.

On the other hand, the level of slips of $L1$ phase measurements
at the fundamental GPS frequency (see Table 3 and Figure~5, at
the right) has a clearly sporadic character, is virtually
independent of the type of receivers, the geomagnetic activity
level, and of the time of day, and is at least one order of
magnitude lower than that in $L1-L2$ measurements. This leads us
to conjecture that the slips of $L1-L2$ measurements are most
likely to be caused by the high level of slips of $L2$ phase
measurements at the auxiliary frequency. According to our data,
these slips are observed at equatorial latitudes under quiet
conditions as well, and at mid-latitudes they increase with
increasing geomagnetic activity.

\section{Discussion and Conclusions}

The main results of this study may be summarized as follows:

\begin{enumerate}

\item We have detected a dependence of the relative density of
phase slips in the navigation system GPS on the disturbance level
of the Earth's magnetosphere during major magnetic storms. Hence,
not only can the disturbances in the near-terrestrial space
environment (caused by corresponding processes in the
\char'134Sun-Earth" system) be detected in TEC measurements by
processing the GPS data, as has now been demonstrated in a large
number of studies, but they also affect the operation of the
navigation system GPS itself.

\item During major magnetic storms the relative density of phase
slips at mid-latitudes exceeds that for magnetically quiet days
at least by one or two orders of magnitude, and reaches a few and
(for some of the GPS satellites) even ten percent of the total
observation density.

\item The level of phase slips for the GPS satellites on the
sunward side of the Earth is by a factor of 5--10 times than that
on the opposite side of the Earth.

\item
The high positive correlation of an increase in the density of
phase slips $P(t)$ and the intensity of ionospheric
irregularities $A(t)$ during geomagnetic disturbances as detected
in this study points to the fact that the increase is slips
$P(t)$ is caused by the scattering of the GPS signal from
ionospheric irregularities.

It is most likely that our recorded phase slips at mid-latitudes
are due to a strong scattering of satellite signals from
small-scale ionospheric $F_2$-layer irregularities which are most
frequently observed at equatorial and polar latitudes [{\it
Aarons et al}., 1996, 1997; {\it Klobuchar}, 1997; {\it Pi et
al}., 1997; {\it Aarons and Lin}, 1999].

This results in a decrease of the "signal/noise" ratio, which has a
particularly strong effect when receiving the signal of the auxiliary
frequency $f_2$, whose power at the GPS satellite transmitter output is
an order of magnitude smaller compared with the fundamental
frequency $f_1$ ( [{\it Langley}, 1998]; Interface Control Document
ICD 200c; http://www.navcen.uscg.mil/pubs/gps/icd200/). Different
types of GPS receivers respond to this differently; on the whole,
however, the picture of the dependence on the local time, latitude
range, and on the level of geomagnetic activity remains sufficiently
stable.

\end{enumerate}

Our intention was to investigate how magnetospheric disturbances
accompanying magnetic storms affect the operation of the GPS
system. A detailed analysis of the factors responsible for the
phase slips in the GPS system is a highly difficult task, and is
beyond the scope of this paper.

Of course, phase measurements are more sensitive to equipment
failures and to interferences in the GPS
\char'134satellite-receiver" channel when compared with group
delay measurements which are directly used in solving navigation
problems. Therefore, it is necessary to have a monitoring of the
errors of determining the coordinates of stationary sites of the
global GPS network, based on the data in the RINEX-format
available from the Internet, and to analyze these series in
conjunction with the data on the conditions of the
near-terrestrial space environment.

We are aware that this study has revealed only the key averaged
patterns of this influence, and we hope that it would give
impetus to a wide variety of more detailed investigations.

\acknowledgments{ Authors are grateful to V.~G.~Eselevich,
V.~V.~Yevstafiev, and G.~V.~Popov for their encouraging interest
in this study and active participations in discussions. We are
also indebted to S.~A.~Nechaev for the data from magnetic
observatory Irkutsk made available to us. Thanks are also due
V.~G.~Mikhalkovsky for his assistance in preparing the English
version of the \TeX-manuscript. Finally, the author wish to thank
the referees for valuable suggestions which greatly improved the
presentation of this paper. This work was done with support under
RFBR grant of leading scientific schools of the Russian
Federation No. 00-15-98509 and Russian Foundation for Basic
Research (grant 99-05-64753).}

{}
\end{article}
\end{document}